\documentclass{article}
\title{Large center vortices and confinement in $3D$ $\Z_2$ gauge theory.}
\author{F.Gliozzi$^a$, M.Panero$^a$ and P. Provero$^{b,a}$}
\date{April 2002}
\usepackage{amssymb}
\usepackage{graphicx}
\usepackage{latexsym}
\newcommand{\eq}{\begin{equation}}
\newcommand{\qe}{\end{equation}}
\newcommand{\ear}{\begin{eqnarray}}
\newcommand{\rae}{\end{eqnarray}}
\newcommand{\Z}{\mathbb{Z}}
\newcommand{\bra}{\langle}
\newcommand{\ket}{\rangle}
\begin{document}
\maketitle
\noindent
$^a$ Dipartimento di Fisica Teorica, Universit\`a di Torino and INFN,
sezione di Torino, via P. Giuria, 1, I-10125 Torino, Italy.
\vskip0.5cm\noindent
$^b$ Dipartimento di Scienze e Tecnologie Avanzate, Universit\`a del
Piemonte Orientale ``A. Avogadro'', and INFN, gruppo collegato di
Alessandria, I-15100 Alessandria, Italy.
\vskip0.5cm\noindent
e-mail: gliozzi, panero, provero@to.infn.it
\begin{abstract}
\noindent
We study the role of large  clusters of center vortices in
producing confinement in $3D$ $\Z_2$ gauge theory. First, we modify each
configuration of a Monte Carlo-generated ensemble in the confined
phase by removing the largest cluster of center vortices, and show that the
ensemble thus obtained does not confine. Conversely, we show that
removing all of the small clusters of center vortices and leaving the
largest one 
only, confinement is preserved, albeit with a string tension
significantly smaller than the original one. Remarkably, also the
string corrections due to the quantum fluctuations of the confining
flux tube are preserved by this transformation.
\end{abstract}
It is widely believed that center vortices play an important role in
producing confinement, by disordering the gauge configurations and hence 
making the Wilson loop decay with the area law. 
The idea itself is quite old \cite{'tHooft:1977hy,Mack:1980rc}, but
its investigation by lattice methods was initiated a few years ago in
Ref. \cite{DelDebbio:1996mh}. A very recent review can be found in
\cite{Faber:2002ib}. 
\par
$3D$ $\Z_2$ gauge
theory is the simplest non trivial gauge theory to display
confinement: due to the moderate size of its configuration space, it
lends itself to very high precision Monte Carlo simulations. Moreover
the gauge group coincides with its center, so that the identification of
center vortices can be performed without resorting to any gauge-fixing
procedure.  
Therefore $\Z_2$ gauge theory appears as an ideal laboratory to test
the actual relevance of center vortices to the mechanism of
confinement.
\par
On the other hand, the role of center vortices in confinement in this
model is in a way trivial, since removal of {\em all} center vortices
from $\Z_2$ gauge theory configurations simply removes all the
dynamics by transforming every configuration into the trivial
vacuum. Therefore center vortices are, in this sense, trivially responsible not only
for confinement, but for the whole dynamics of the model.
\par
In this study we investigate a subtler issue, namely the effect of the
{\em size} of clusters of center vortices on confinement. The hypothesis we
want to test, first proposed in  Ref.\cite{Engelhardt:1999fd}, is that
confinement in this model is due to the 
existence of an infinite cluster of center vortices, that is a
connected 
component, in the graph defined in the dual lattice by all the
center vortices, whose size scales linearly with the lattice
volume. Only such a giant component in the graph is able to disorder
the gauge configurations 
enough to produce the area-law decay of the Wilson loop.
\footnote{In the 3D $\Z_2$ gauge theory we are studying there exist
also a {\it different} kind of cluster whose percolation is associated to
confinement: the Fortuin-Kasteleyn clusters defined in the dual spin
model \cite{nato}. See also Ref.~\cite{Gliozzi:2001tu} for an 
analysis of the relationships between these and other condensates in
the case of
$\Z_2$ gauge theory with dynamical matter.}
\par
The $\Z_2$ gauge model is defined by the action
\eq
S(\beta)=-\beta\sum_{\Box}\sigma_{\hskip -.15 cm 
\matrix{~\\ \Box}}  ~~~~,~
\sigma_{\hskip -.15 cm \matrix{~\\\Box}}=\prod_{\ell\in\,\Box}\sigma_{\ell}
\qe
where the sum is extended to all plaquettes of a cubic lattice, on
whose links $\ell$ $\Z_2$ variables $\sigma$ are defined: each plaquette
contributes the product of its links to the action.
Center vortices are constructed by assigning a vortex in the dual
lattice to each frustrated plaquette in the direct lattice. Since, in
the direct lattice, the product of the six plaquettes forming a cube
is constrained to be equal to 1, the resulting graph of center
vortices in the dual graph has even coordination number. 
\par
Such graph is in general made of many connected components, that we
call {\em clusters} of center vortices. The value of a Wilson loop in
a given configuration is $\pm 1$ according to the number, modulo 2, of
frustrated plaquettes of an arbitrary surface bounded by the loop, or, 
in the center vortex
language, to the number, modulo 2, of vortex lines that are linked to
the loop. 
\par
A plausibility argument for the crucial role of the infinite cluster
of vortex lines goes as follows (see also Ref. \cite{Engelhardt:1999fd}): 
first, note that the only clusters contributing to the Wilson loop
$W(C)$ are those linked to $C$, as the property of even coordination
number requires.
Assume now that the clusters of center vortices have a maximum size of order
$R$: the contribution to a Wilson loop of much larger size comes only
from the clusters located near the loop $C$.  The number of these
clusters grows linearly with the length of the loop and produces the
decay of $\bra W(C)\ket$ with the perimeter law and the theory is
deconfined. If such a
maximum size $R$ does not exist, that is if there is an infinite 
cluster, then the number of vortex lines linked to the loop grows with
its area, and we have confinement.
\par
It is straightforward to verify numerically that in the confining phase 
of $\Z_2$ gauge system there is no ambiguity in finding  a cluster of 
center vortices whose size scales linearly with the lattice volume for 
large enough lattices (see Fig.1), while in the deconfined phase we
found  that the density of the largest cluster decreases rapidly with
the volume.
   
It has to be noted that the presence of the infinite cluster does not 
necessarily imply a percolation property of the central
vortices. Confinement requires merely the presence of an infinite
cluster, while  percolation demands an infinite cluster of large
enough density.
\par
\begin{figure}
\centering{\includegraphics[width=11.cm]{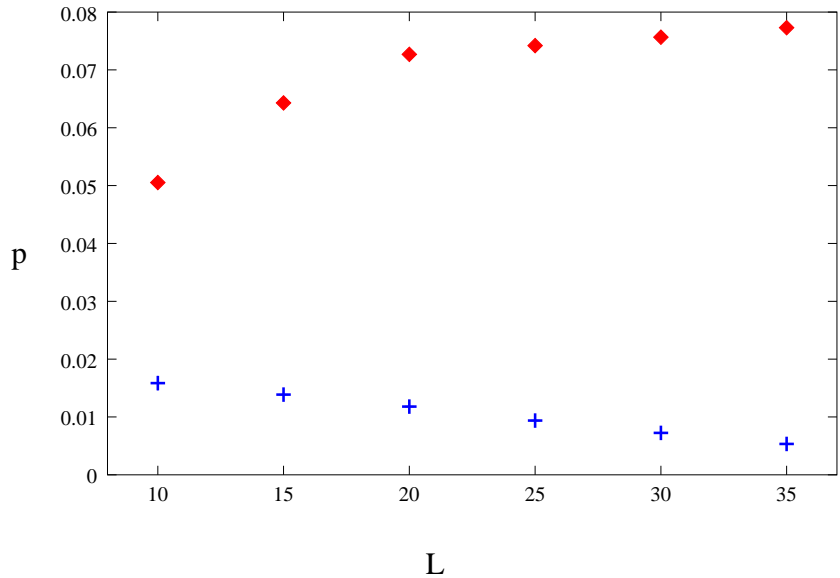}}
\caption{\sl The diamonds represent the density $p$ of dual links
belonging to the largest cluster, as a function of the lattice size:
the fact that such density tends to 
a constant means that the largest cluster of center vortices is
infinite in th thermodynamic limit. The same density for the second 
largest cluster tends to zero (crosses)}
\end{figure}
To study more carefully the relationship between the presence of the 
infinite cluster and the value string tension,
we
 chose to simulate the model at $\beta=0.74883$, which is well
inside the scaling region, and for which the value of the string
tension is known with high precision from simulations of the dual
model, that is the 3D (spin) Ising model 
\cite{Hasenbusch:1992zz,Caselle:1994df}
\eq
\sigma=0.01473(10)
\qe
\par
First, we verified the {\em existence}
of an ``infinite'' cluster of vortex lines: for each configuration, we
selected the largest connected component of the graph defined by the
center vortices, and verified that the size of such component grows
linearly with the volume of the lattice.
\par 
The results of this analysis are shown in Fig. 1, where the size of
the largest cluster divided by the lattice volume is shown to approach
a constant for large lattices. The size of the second largest cluster
is also shown: its size relative to the lattice volume tends to
zero and the identification of the ``infinite'' cluster is
unambiguous.
To test the relevance of the largest cluster of center vortices,
we proceeded as follows:
first, we modified each configuration in the Monte Carlo ensemble 
by eliminating  all the vortices not belonging to the largest
cluster, and second, by eliminating, instead, the largest cluster
only. The qualitative picture described above suggests that a non-zero
string tension will be found in the first case but not in the second.
\par
\par
An efficient method to extract the string tension from Wilson loop
data generated by Monte Carlo simulations, which takes into account
the string fluctuation contribution, was introduced in
Ref.~\cite{Caselle:1996ii}. One
defines the ratio of the expectation values of rectangular Wilson
loops with the same perimeter:
\eq 
r(L,n)=\frac{\langle W(L+n,L-n)\rangle}{\langle W(L,L)\rangle}
\qe
If a simple area law described the Wilson loop behavior, such ratios
would behave for large $L$ as
\eq
r(L,n)\sim \exp\left({\sigma n^2}\right)
\qe
The quantum fluctuations of the color flux tube induce a correction
to this behavior that depends on the ratio $n/L$ only:
\eq
r(L,n)\sim \exp\left({\sigma n^2}\right)\  F\left(n/L\right)
\qe
where 
\eq
F(t)=\left[\frac{\eta(i)\sqrt{1-t}}{\eta\left(i\frac{1+t}{1-t}\right)}\right]^{1/2}
\label{f}
\qe
and $\eta$ is the Dedekind function, 
so that one can
define the modified ratios
\eq 
s(L,n)=\frac{r(L,n)}{F\left(n/L\right)}
\qe
which in turn can be used to evaluate the string tension, since the quantity
\eq
\sigma_{\rm eff}(L,n)\equiv\frac{\log s(L,n)}{n^2}
\label{sigma}
\qe
approaches the string tension for large $L$.
\par
A test of the correctness of the description Eq.~(\ref{f}) 
of the string fluctuation is obtained by verifying that the estimator
$\sigma_{\rm eff}$ defined in 
Eq.~(\ref{sigma}) reaches a constant value for smaller values of $L$
than the simple estimator which neglects string fluctuations:
\eq
\sigma_{\rm no\ string}\equiv \frac{\log r(L,n)}{n^2} 
\label{sigma_nostring}
\qe
\par
In Fig. 2 we display the behavior of $\sigma_{\rm eff}(L,n)$ as a
function of $L$ for $n=2,3,4$ for the original configurations. Fig. 3
shows the same for the configurations where only the largest
cluster of center vortices has been left, while Fig. 4 is obtained
by removing only the largest cluster.
\begin{figure}
\centering{\includegraphics[width=12.cm]{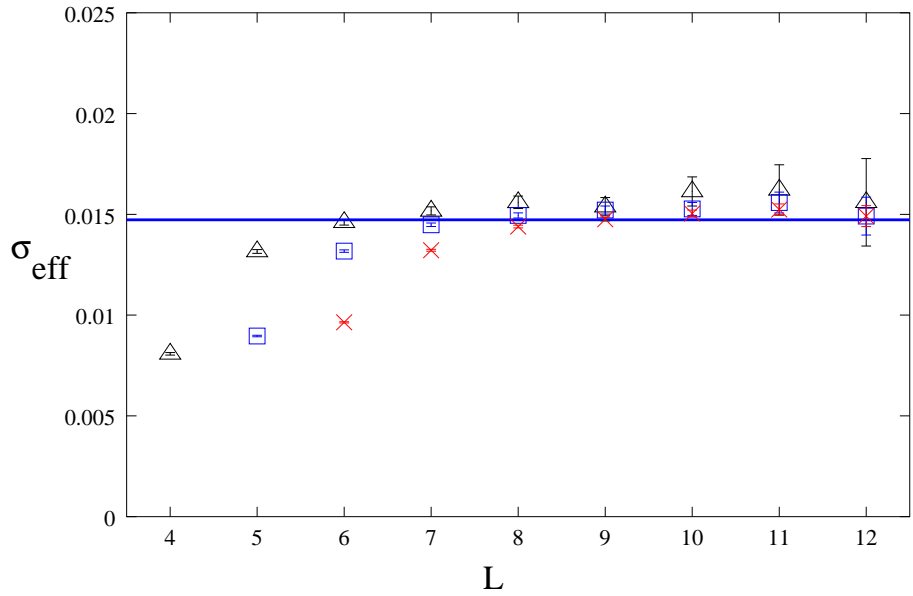}}
\caption{\sl The estimator $\sigma_{\rm eff}$ of the string tension
for $n=2$ (triangles), 3 (squares) and 4 (crosses), as a function of
$L$. The horizontal line is the string tension $\sigma=0.1473$, taken
from the literature.}
\end{figure}
\begin{figure}
\centering{\includegraphics[width=12.cm]{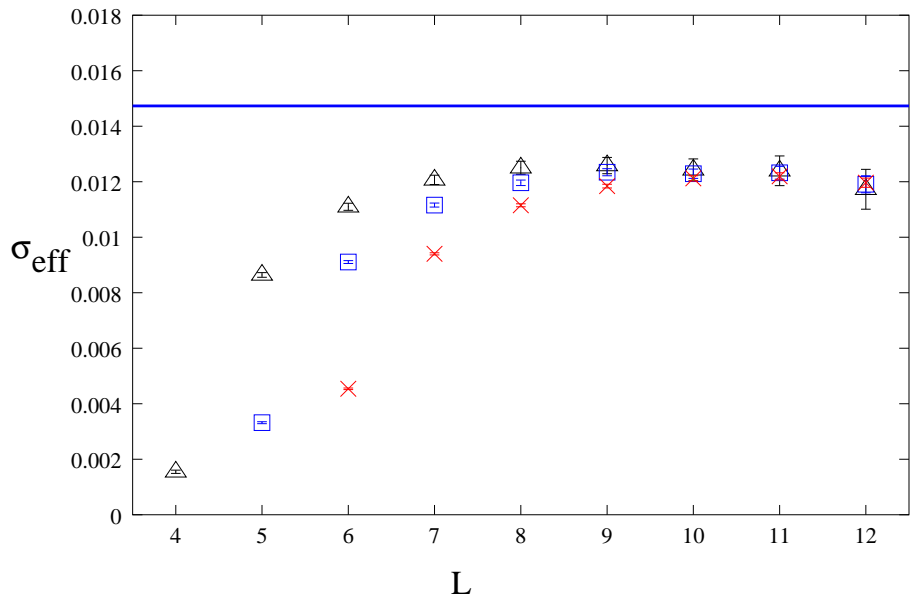}}
\caption{\sl Same a s Fig. 2 for the configurations where all of the
vortices except the ones belonging to the largest cluster have
been removed.}
\end{figure}
\begin{figure}
\centering{\includegraphics[width=12.cm]{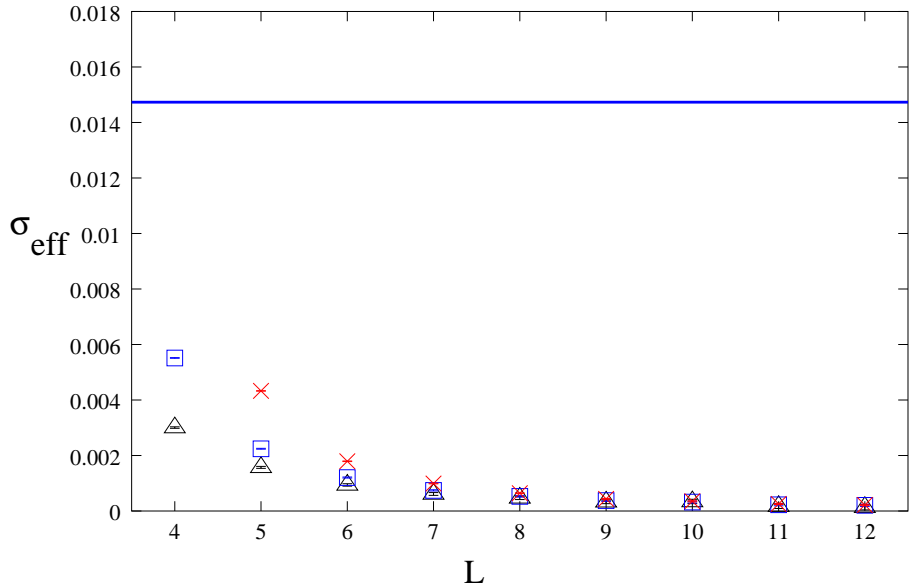}}
\caption{\sl Same as Fig. 2 for the configurations where the vortices
belonging to the largest cluster have
been removed.}
\end{figure}
\par
These figures show that our main expectations are fulfilled: by
removing the finite clusters one still obtains a confining
theory (Fig. 3), while when removing the ``infinite'' cluster the
string tension vanishes (Fig. 4).
\par
However by examining Fig. 3 one can learn two non-trivial lessons:
first, the string tension of the ensemble where only the largest
cluster has been retained is definitely smaller than the one of the
full model: while it is true (as shown by Fig. 4) that finite
clusters cannot by themselves induce confinement, it is not true that
they do not contribute to the string tension.
\footnote{Note that the qualitative argument given above would imply that the
string tension generated by the largest cluster should equal the
full one {\it only} by assuming that the space-time distributions of
the infinite and finite clusters are uncorrelated, an
assumption which is obviously unjustified in any non-trivial theory.}
\par
Second, Fig. 3 shows that the corrections due to the flux tube
fluctuations survive the elimination of the small clusters: indeed
Fig. 3 shows the estimator $\sigma_{\rm eff}$ defined in
Eq.~(\ref{sigma}) which includes the string fluctuation contribution.
The quantitative importance of such contributions is shown by
comparing Fig. 3 to Fig. 5, where the two string tension estimators
$\sigma_{\rm eff}$ and $\sigma_{\rm no\ string}$ are compared for the
$n=4$ data in the configurations in which only the largest  cluster
has been left.
\begin{figure}
\centering{\includegraphics[width=11.cm]{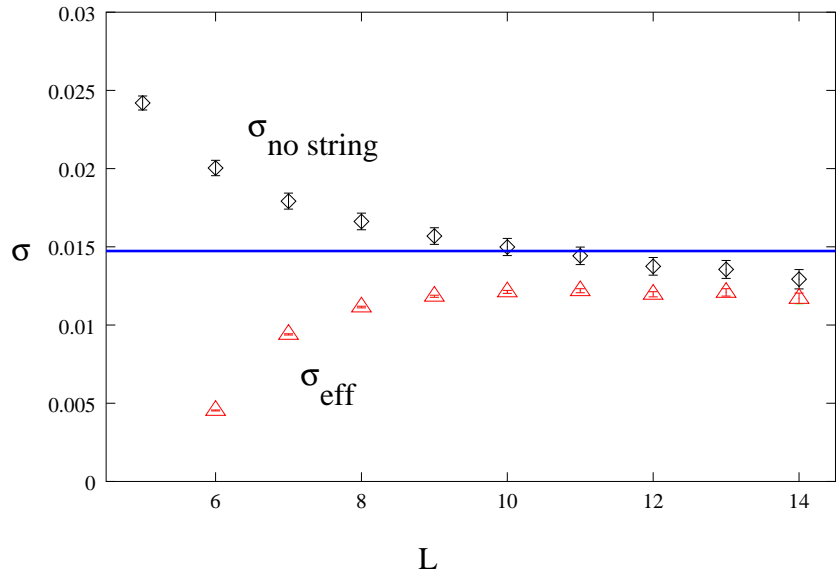}}
\caption{\sl Comparison between the two estimators $\sigma_{eff}(L,n)$ and
$\sigma_{\rm no\ string}(L,n)$. The data refer to $n=4$ and to the
configurations where all but the largest cluster have been removed.}
\end{figure}
\vskip1.cm
In conclusion, our results confirm the picture of confinement as
due to the existence of an infinite cluster of center vortices: our
choice of the $\Z_2$ gauge theory allows us to bypass all the problems
related to the gauge-fixing and center projection that one encounters
when studying the same issue in $SU(N)$ gauge theories.
Two important new facts emerge from our study:
\begin{itemize}
\item
While the largest center vortex is responsible for confinement,
since its removal from the configurations makes the string tension
vanish, the string tension measured from configurations in which all
the other clusters have been removed does not reproduce the full
string tension of the original theory. Therefore small clusters of
vortices, while unable by themselves to disorder the system enough to
produce confinement, do give a finite contribution to the string
tension of the full theory.
\item
The quantum fluctuations of the flux tube survive the elimination of
the small clusters: the Wilson loop after deletion of all the
small clusters show the same shape dependence as the ones of
the full theory, which can be explained as originating by the
fluctuations of a free bosonic string.
\end{itemize}

\end{document}